\documentclass[epj]{mysvjour}
\usepackage{latexsym}
\usepackage{graphics}
\usepackage{a4wide}
\usepackage{amsmath}
\usepackage{bbm}
\usepackage{amsfonts}
\usepackage{cite}
\usepackage{enumerate}

\begin{document}

\title{Minimizing driver's irritation at a roadblock}
\author{C.J.J. Vleugels\thanks{\emph{e-mail:} c.j.j.vleugels@gmail.com}, A. Muntean\inst{1} \inst{2} \and M.J.H. Anthonissen\inst{1} \and T.I. Seidman\inst{3}}

\institute{
Centre for Analysis, Scientific computing and Applications, Department of Mathematics and Computer Science, Eindhoven University of Technology, PO Box 513, 5600 \ MB Eindhoven, The Netherlands
\and Institute for Complex Molecular Systems, Eindhoven University of Technology, PO Box 513, 5600MB Eindhoven, The Netherlands
\and Department of Mathematics and Statistics, University of Maryland Baltimore County, MD 21250 Baltimore, USA
}

\date{Received: date / Revised version: date}
%
\abstract{
Urban traffic is  a logistic issue which can have many societal implications, especially when,  due to a too high density of cars, the network of streets of a city becomes blocked, and consequently,  pedestrians, bicycles,  and cars start sharing the same traffic conditions potentially leading to high irritations (of people) and therefore to chaos.
In this paper we focus our attention on a simple scenario: We model the driver's irritation induced by the presence of a roadblock. As a natural generalization, we extend the model for the  two one-way crossroads traffic
presented in \cite{OPTI01} to that of a roadblock.
Our discrete model defines and minimizes the total waiting time. The novelty lies in introducing the (total)
driver's irritation and its minimization.
Finally, we apply our model to a real-world situation: rush hour traffic in Hillegom, The Netherlands. We observe that minimizing the total waiting time and minimizing the total driver's irritation lead to different traffic light strategies.
%
}

\titlerunning{Minimization of irritation at a roadblock}
\authorrunning{C.J.J. Vleugels}
\maketitle

\section{Introduction}\label{intro}
Traffic flow problems are a major topic of research. Examples include traffic on highways or within cities (crossings, traffic light controls, roadblocks and so on).  Many practical questions are still in search for an optimal answer: Which traffic light settings lead to a minimal waiting time per car? Can one control this in an {\em a priori} way? What about the waiting time for all the cars together stacked in a lane? At which speed do incoming vehicles have to drive to maximize the flow during a traffic jam? How can local car-car and car-traffic light interactions lead to emergent coordination of patterns like ``green-waves''?  Most of these questions can be addressed for  not too complex road networks (like simple highway systems)  but are mathematically NP complete (and therefore untractable) at the level of cities with thousands of traffic lights.

In this paper, we propose a somewhat different view and consider the driver's perspective. We develop a model for the driver's irritation\footnote{There are all sorts of causes for irritation in traffic, often actions of other drivers like tailgating, lack of flashing when switching lanes, driving too slow or too fast and so on. It may also be caused by having to wait for a long time at traffic lights or in jams. Here we start from the assumption that high irritations lead with an increased probability  to sudden dangerous motions and herewith to accidents.\\} and propose to minimize this instead of minimizing the total waiting time (TWT). We will see that such a strategy may lead to slightly different clearance policies than when minimizing the TWT. In next papers, we intend  to apply this modeling framework to optimizing for load levels within capacity and will study the dependence of optimization on modeling the concept of ``driver's irritation''.

Within this paper we study one specific traffic situation: a two-way street roadblock where one lane is blocked.
Our model is a generalization of that for the intersection of two urban streets presented in \cite{OPTI01}. Note that there are standard measures taken by the authorities for a roadblock situation. An example is a bypass: traffic is redirected so that there is less or no traffic that has to pass the roadblock. Also, roadwork can be planned during time intervals when there is not much traffic, such as weekends or nights. Sometimes these measures are not possible and one has to look for alternatives such as lowering the irritation with the help of smart traffic light settings. Typically traffic lights are installed such that the TWT is minimized. This means that the total flow of traffic is optimized. 
\\
This work is based on \cite{OPTI01}. More research on traffic flow is done by many researchers, for instance \cite{abc} connects the current research trends in traffic flow to discrete crowd dynamics. Another approach is queueing theory, see e.g. \cite{REF02,PHD01}. We use a discrete deterministic particle systems-like model \cite{OPTI02}. A more stochastic approach of traffic flow problems is found in \cite{BOOK01} and references cited therein. Finally traffic flow can be modeled by fluid-dynamics principles using ordinary and partial differential equations; see,  for instance,  Refs. \cite{SEL03,Fermo2}.\\
\\
In Section~\ref{sec:2}, we present our roadblock model based on \cite{OPTI01}. We minimize the total waiting time. Here we introduce our concept of driver's irritation and study traffic light settings that minimize total irritation. In Section~\ref{sec:3}, we apply our methodology to a real traffic situation, using empirical data for rush hour scenario in Hillegom, The Netherlands. Finally, we conclude with a discussion in Section~\ref{sec:4}.

\section{Roadblock model}\label{sec:2}
\begin{figure}[h!]
\resizebox{\columnwidth}{!}{\includegraphics{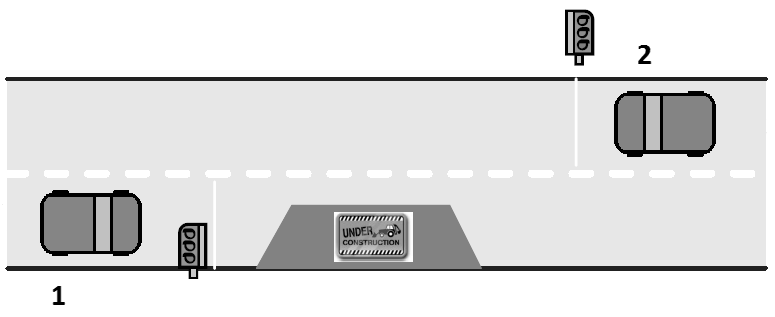}}
\caption{Schematic picture of the roadblock}
\label{fig:1}
\end{figure}

Consider a two-way street where one of the two lanes is blocked. See Figure \ref{fig:1} for a schematic picture of the roadblock. Traffic lights are installed such that both directions do not have green light at the same time. After each time a direction has had green light, both lights are red for a specific time. This is called a transit period and is necessary to clear the road. This leads to the following cycle:
\begin{itemize}
\item	Phase I: direction 1 green: period $T_1$
\item	Phase II: both directions red: period $\tau$
\item	Phase III: direction 2 green: period $T- T_1$
\item	Phase IV: both directions red: period $\tau$
\end{itemize}
After Phase IV, the cycle repeats itself and starts with Phase I again. We take the values for $T$ and $\tau$ constant, and the value for $T_1$ variable. We are now looking for a value for $T_1$ that minimizes the total waiting time.
We note that we can see this problem as a queuing system with two queues and one server, where at every moment no more than one queue is served.

\subsection{Notation. Basic assumptions}\label{sec:2.1}
\begin{itemize}
\item	The arrival rates $\alpha_1, \alpha_2$ [cars/second] and the passing rates $\beta_1, \beta_2$ [cars/second] are assumed to be constant;
\item	The cycle number is given by $n$, and represents in which cycle we are;
\item	The queue lengths of each direction at the beginning are $0$. At time $t$ the queue lengths are: $N_1(t), N_2(t)$;
\item	The total waiting time TWT is defined as the sum of the waiting times of each direction. This waiting time is defined as the product of the total number of cars that have to wait and the time that they have to wait (times of the red phases);
\item	We only consider the \emph{heavy traffic scenario} where during one cycle more cars arrive than leave in both directions. This assumption is made because this scenario is the most interesting and problematic one for drivers, logistics companies and governmental and environmental institutes. Note however, that in this scenario the queue of waiting cars would grow infinitely. In reality, the heavy traffic scenario holds only for limited time (such as during rush hour), which ensures the queue to be bounded.
\end{itemize}
Here we assume the cycle period $T$ is constant. This is a fair assumption for the rush hour case. Choosing $T_1$ fixes $T_2$, the cycle period $T$ playing the role of a parameter.\\

\subsection{Total waiting time}\label{sec:2.2}
Based on these assumptions, we can define the total waiting times for Directions 1 and 2 as:
\begin{equation}
T_{n}^{(w,1)}(T_1) := N_1(nT+T_1)(T-T_1) + \frac{1}{2}\alpha_1(T-T_1)^2.
\end{equation}
\begin{equation}
\begin{split}
T_{n}^{(w,2)}(T_1) := &  N_2(nT)(T_1 + \tau) + N_2((n+1)T-\tau)\tau \\
 & + \frac{1}{2}\alpha_2(T_1 +2\tau)^2.
\end{split}
\end{equation}
The assumption of heavy traffic means that the queue lengths after every phase increase linearly over the cycle number. For example the queue length of Direction 1 is:
\begin{equation}
        N_1(nT+T_1) := n(\alpha_1 T -\beta_1 T_1) + (\alpha_1- \beta_1) T_1.
\end{equation}
Let us substitute these queue lengths into (2). Now if we take the sum of those two waiting times, we get the \emph{total waiting time}, that is the waiting time of all the cars at the roadblock. Minimizing the total waiting time with respect to $T_1$ gives the value:
\begin{equation}\label{eq:T_1}
        T_1 = \frac{(\alpha_1 - \alpha_2 +\beta_1 + \beta_2) T - 4\beta_2 \tau}{2 \beta_1 + 2 \beta_2}.
\end{equation}
We note that if we take $\tau=0$, we get the same value as the one obtained in \cite{OPTI01}, which suggests that the roadblock problem and the two one-way crossroads have the same mathematical structure.

\subsection{Driver's irritation}\label{sec:2.3}

Instead of minimizing the total waiting time, we can take a different look at the traffic situation. We like now to think in terms of minimizing the driver's irritation. First of all, we have to define our concept of irritation. We say that the total irritation per cycle of one direction depends on the irritation per car. This irritation per car is defined at the moment that the lights switch from green to red. We call this the \emph{crucial moment} because at that time there is a significant change in the irritation: it increases because the people have to wait while they first were able to pass the roadblock. We note that the crucial moment for Direction 1 is after Phase I and for Direction 2 after Phase III. In this framework, we assume the change in the irritation per car depends on several causes:
\begin{enumerate}[(i)]
\item	\emph{Waiting time}; the longer you have to wait after the crucial moment, the higher the irritation;
\item	\emph{Queue length of the other direction}; it is irritating when you have to wait in front of a red light, but it is even more irritating when there are no (or few) cars passing from the other direction. So the queue lengths of the other direction and the irritation per car are inversely proportional;
\item	\emph{The distance of the car to the traffic light}; If we assume that the distances between two cars in the queue is constant, the distance to the traffic light can be written in terms of the position $k$ of the car in the queue. We have in mind here two conceptually different scenarios:
    \begin{itemize}
        \item	The closer you are to the traffic light, the higher the irritation because if the light had been green a little longer, you were able to pass the roadblock (we call this \textbf{Case I}). In this case, the position $k$ in the queue and the irritation per car are inversely proportional. We use the function $f(k) = {C_{I}}/{k}$, where $C_{I} \geq 0$ is a dimensionless constant (hence $f(k)$ is dimensionless);
        \item	The further away you are from the traffic light at the crucial moment, the higher your irritation (there are still so many cars in front of you). We call this \textbf{Case II}. In this case, the position $k$ in the queue and the irritation per car are directly proportional. We use the function $f(k) = C_{II} k$, where $C_{II} \geq 0$ is a dimensionless constant (hence $f(k)$ is dimensionless).
    \end{itemize}
\end{enumerate}
Assuming that (i), (ii) and (iii) affect irritation independently, gives us the following expressions for the irritation per car in Direction 1 ($i^{(1)}(k)$) and 2 ($i^{(2)}(k)$):
\begin{equation}\label{ir5}
\begin{array}{l l}
     i^{(1)}(k,T_1) & := \dfrac{f(k)}{N_{2}(nT+T_1)+1}(T-T_1),\\ \\
     i^{(2)}(k,T_1) & := \dfrac{f(k)}{N_{1}((n+1)T-\tau)+1}(T_1 + 2\tau).\\
\end{array}
\end{equation}
This irritation per car of each direction has units $s \cdot car^{-1}$ because $f(k)$ is dimensionless, $N_{j}$ has dimension $car$ and $T,T_1$ have dimension $seconds$ for $j=1,2$. Taking the sum of the irritation per car (in both directions) defines the \textit{total irritation} at the crucial moment. But for each direction, there is only one crucial moment. At all the other moments, the light stays red or switches from red to green. At these moments, we assume that there is no extra irritation per car, only the irritation that arises from the cars that have to wait because they have arrived during a red phase. This gives the following expression:
\small
\begin{equation}\label{eq:irdir1}
I_{n}^{(1)}(T_1) := \sum_{k=1}^{N_{1}(nT+T_1)} i^{(1)}(k,T_1) + \frac{C_{1}}{2}\alpha_1 (T-T_1)^{2},
\end{equation}
\normalsize
which is the sum of the irritation per car of direction 1 at time $nT+T_1$ plus the irritation that occurs from the waiting time of cars that arrive in the red phase which lasted $T-T_1$ seconds. Note that it is needed to multiply the second term with a constant $C_{1}$ (with units $car^{-2}$) to make sure that the terms entering (6) (and (7)) have the same units. Similarly, we have:
\small
\begin{equation}\label{eq:irdir2}
I_{n}^{(2)}(T_1) := \sum_{k=1}^{N_{2}((n+1)T-\tau)} i^{(2)}(k,T_1) + \frac{C_{2}}{2}\alpha_2 (T_1+2\tau)^{2},
\end{equation}
\normalsize
which is the sum of the irritation per car over all the cars in queue 2 plus the irritation that occurs from the waiting time of cars that arrive during the red phases of direction 2. Again, $C_{2}$ is a constant with units $car^{-2}$ to make sure all the terms in (6) have the same units. The total irritation in cycle $n$ is then defined as the sum of \eqref{eq:irdir1} and \eqref{eq:irdir2}:
\begin{equation}\label{eq:TI}
TI(T_1) := I_{n}^{(1)}(T_1) + I_{n}^{(2)}(T_1)
\end{equation}
This total irritation has units in $s \cdot car^{-1}$. The way the irritation is defined makes it very complicated to minimize this and get a simple analytical expression for $T_1$.

In the next section, we investigate numerically the effect of minimizing the total waiting time and the total irritation $TI$ using real traffic data.

\section{Application: Rush hour in Hillegom}\label{sec:3}

\begin{figure}[h!]
\resizebox{1.0\columnwidth}{!}{\includegraphics{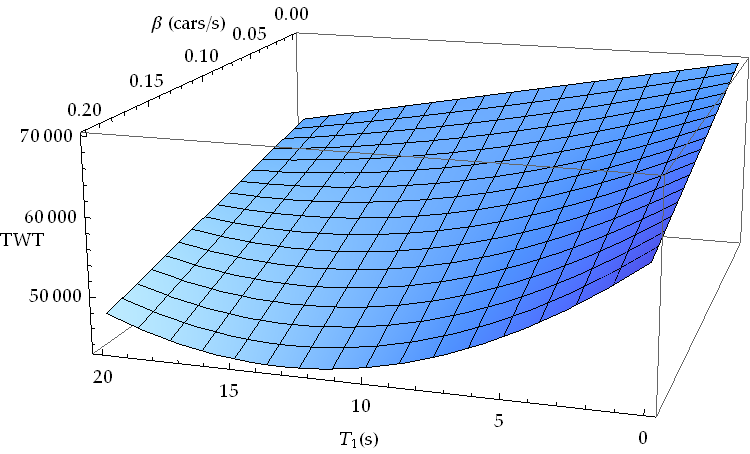}}
\caption{3D plot of the total waiting time against the different green times and passing rates for the Leidsestraat during morning rush hour.}
\label{fig:2}
\end{figure}

\begin{figure}[h!]
\resizebox{1.0\columnwidth}{!}{\includegraphics{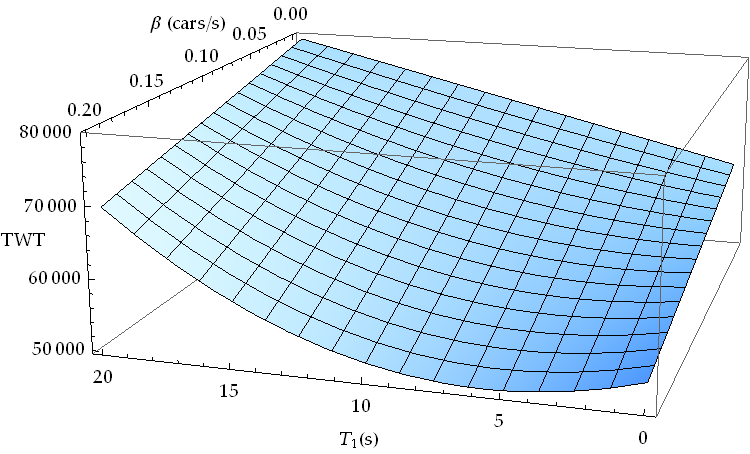}}
\caption{3D plot of the total waiting time against the different green times and passing rates for the Leidsestraat during evening rush hour.}
\label{fig:3}
\end{figure}

\begin{figure}[h!]
\resizebox{1.0\columnwidth}{!}{\includegraphics{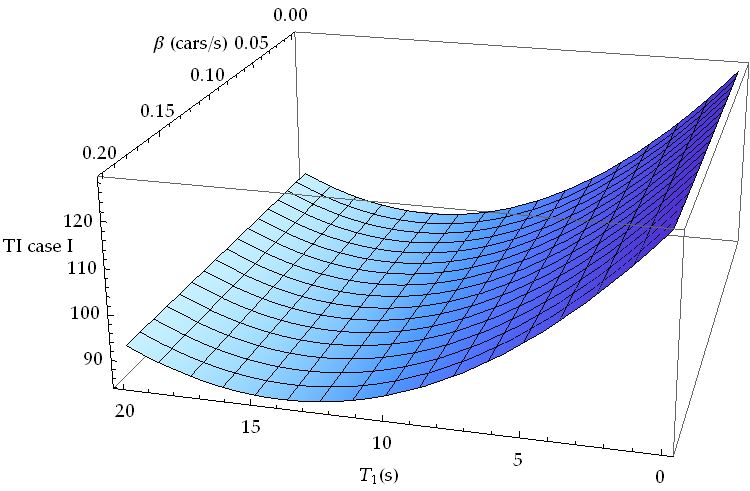}}
\caption{3D plot of the total irritation in Case I against the different green times and passing rates for the Leidsestraat during morning rush hour.}
\label{fig:irImo}
\end{figure}

\begin{figure}[h!]
\resizebox{1.0\columnwidth}{!}{\includegraphics{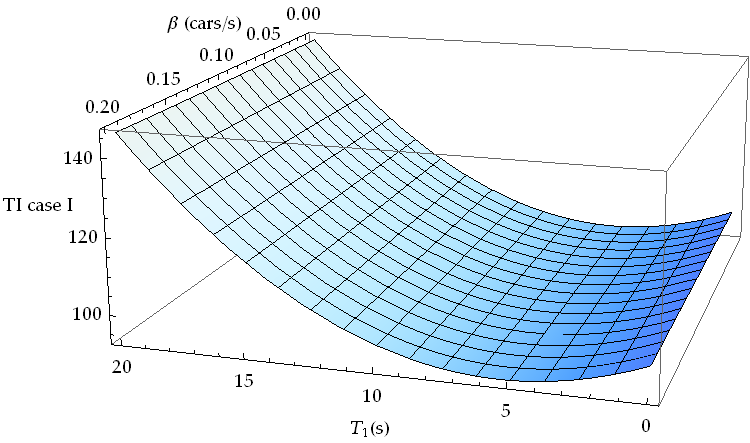}}
\caption{3D plot of the total irritation in Case I against the different green times and passing rates for the Leidsestraat during evening rush hour.}
\label{fig:irIev}
\end{figure}

\begin{figure}[h!]
\resizebox{1.0\columnwidth}{!}{\includegraphics{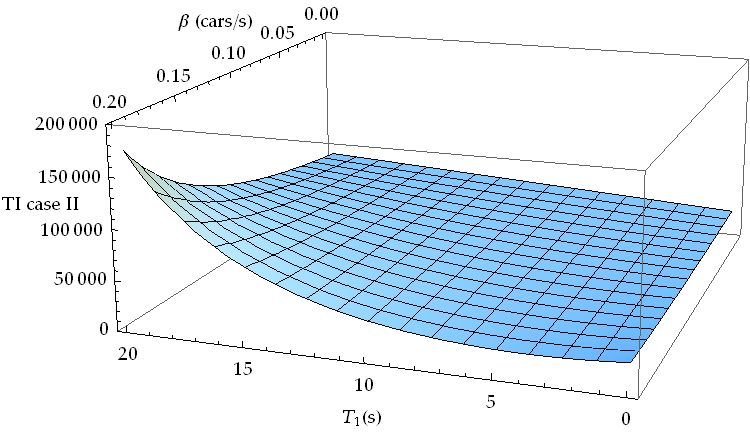}}
\caption{3D plot of the total irritation in Case II against the different green times and passing rates for the Leidsestraat during morning rush hour.}
\label{fig:irIImo}
\end{figure}

\begin{figure}[h!]
\resizebox{1.0\columnwidth}{!}{\includegraphics{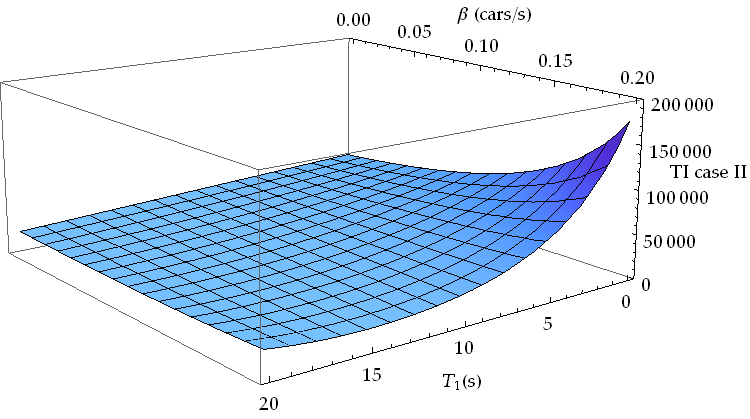}}
\caption{3D plot of the total irritation in Case II against the different green times and passing rates for the Leidsestraat during evening rush hour.}
\label{fig:irIIev}
\end{figure}

\begin{figure}[h!]
\resizebox{\columnwidth}{!}{\includegraphics{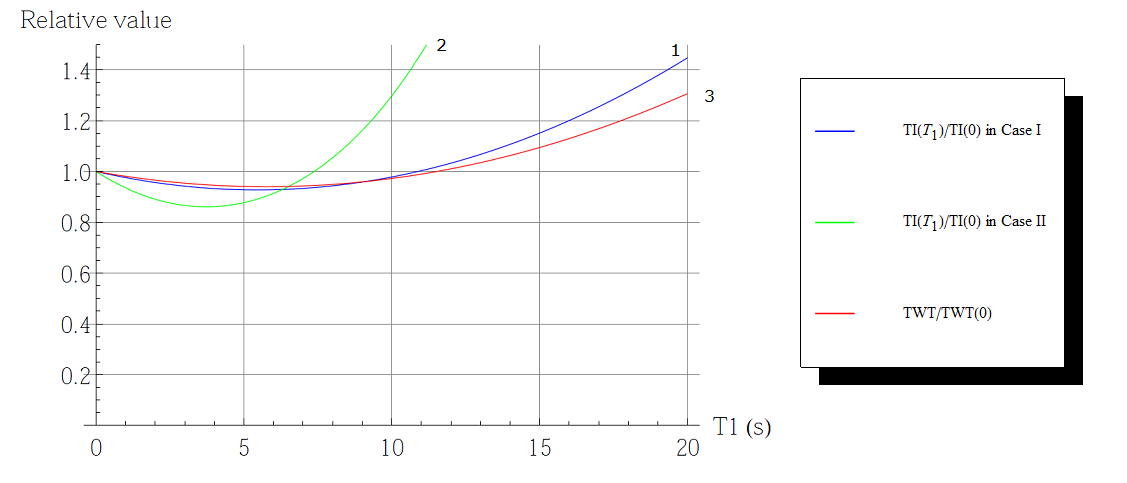}}
\caption{Irritation and total waiting time for Leidsestraat during morning rush hour, where\newline
$1$ is given by $\dfrac{TI(T_1)}{TI(0)}$ in Case I vs. $T_1$,\newline
$2$ is given by $\dfrac{TI(T_1)}{TI(0)}$ in Case II vs. $T_1$ and\newline
$3$ is given by $\dfrac{TWT(T_1)}{TWT(0)}$ vs. $T_1$.
}
\label{fig:4}
\end{figure}

\begin{figure}[h!]
\resizebox{\columnwidth}{!}{\includegraphics{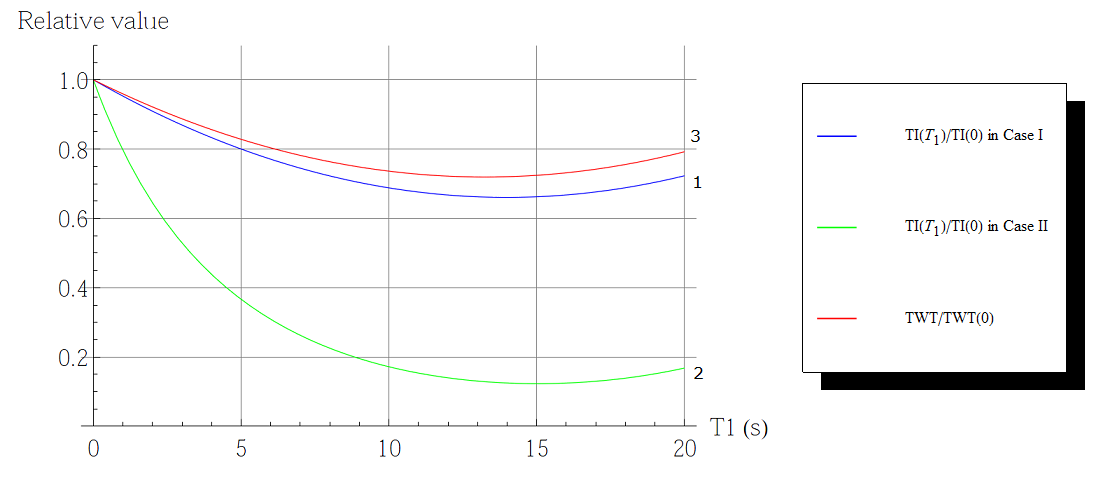}}
\caption{Irritation and total waiting time for Leidsestraat during evening rush hour, where\newline
$1$ is given by $\dfrac{TI(T_1)}{TI(0)}$ in Case I vs. $T_1$,\newline
$2$ is given by $\dfrac{TI(T_1)}{TI(0)}$ in Case II vs. $T_1$ and\newline
$3$ is given by $\dfrac{TWT(T_1)}{TWT(0)}$ vs. $T_1$.
}
\label{fig:5}
\end{figure}

In this section we use the empirical data that we received from Peter Veeke, Traffic and Transportation Advisory Consultant for Breijn B.V. The data contains traffic intensities (in cars per hour) during rush hour for some main streets in Hillegom, The Netherlands. The traffic intensities in cars per hour can be converted into arrival rates in cars per second. Here, we only consider one main street in Hillegom: the Leidsestraat which lies between Singel and Olympia. This part of the Leidsestraat is a two-way street. The corresponding the arrival rates are given in Table 1. We note that during the morning rush hour, direction $N\rightarrow S$ has a larger arrival rate, while in the evening, direction $S\rightarrow N$ has a larger arrival rate.\\
\begin{table}[h!]
\caption{Arrival rates for Leidsestraat}
\label{tab:1}
\begin{center}
\begin{tabular}{ccc}
\hline\noalign{\smallskip}
 Direction & Arrival rate & Arrival rate \\
  & (morning) & (evening)\\
\noalign{\smallskip}\hline\noalign{\smallskip}
$S\rightarrow N$ & $0.190$ & $0.264$\\
$N\rightarrow S$ & $0.302$ & $0.176$\\
\noalign{\smallskip}\hline
\end{tabular}
\end{center}
\end{table}\\
At this point, we assume that there is a roadblock in the Leidsestraat between Singel and Olympia. Given only the arrival rates in Table 1, we cannot determine any values for $T_1$ yet. Therefore, we have to make some assumptions about the corresponding passing rates and about the cycle and transit period. Because the Leidsestraat is a major street and the traffic intensities are high during rush hour, we assume heavy traffic. We take the cycle period equal to 30 seconds, and the transit period equal to 5 seconds. We have not received any data about the passing rates. Consequently, we need to estimate the passing rates. We assume both direction have the same passing rates, resulting in $\beta=\beta_1=\beta_2$. Doing the same type of analysis as in \cite{OPTI01}, we obtain the following inequalities belonging to a heavy traffic scenario:
\begin{equation}\label{s1}
(\alpha_2 + 3 \alpha_1)\beta T - 4\beta^2 \tau > 2 \beta^2 T
\end{equation}
\begin{equation}\label{s2}
(\alpha_1 + 3 \alpha_2)\beta T - 4\beta^2 \tau > 2 \beta^2 T
\end{equation}
Hence, our passing rate is typically varying between $0$ to $0.2$. Larger values would simply mean that the traffic scenario cannot be considered as heavy traffic anymore. Plotting the total waiting time as a function of the value for $T_1$ and the passing rate $\beta=\beta_1=\beta_2$ give us Figure \ref{fig:2} and \ref{fig:3}.\\
\\
Obviously, we see that the largest value for $\beta$ results in the smallest total waiting time. But it is not obvious that the largest value for $\beta$ also results in the smallest irritation, because of the construction of the irritation per car. So, we plot the irritation in case I as a function of $\beta$ and $T_1$ which results in Figure \ref{fig:irImo} for the morning and \ref{fig:irIev} for the evening rush hour scenario. For the irritation in case II we obtain Figure \ref{fig:irIImo} and \ref{fig:irIIev} for morning and evening rush hour, respectively.\\
\\
We see in all figures that $\beta=0.2$ results in a minimal total waiting time. Therefore, we fix the passing rates to $0.2$. We note that if we fix $\beta$ and let $\alpha$ increase, we obtain a larger total waiting time and irritation. This is due to fact that increasing $\alpha$ is the opposite effect of increasing $\beta$. Most likely, cars queues grow unboundently and the traffic enters the unstable regime.
\\
Now, we are able to calculate the values for the green-times $T_1$. Because the total waiting time and the irritation have different dimensions and magnitudes, plotting them against the $T_1$ will not gives us a nice figure. If we plot the total waiting time and the irritation in Case I and II divided by their value at $T_1 =0$, we get functions which start at the point $(0,1)$ that we use to compare the locations of the minima; see Figure \ref{fig:4} and \ref{fig:5}. Figure \ref{fig:4} shows that the total waiting time during morning rush hours is minimal for $T_1 = 5.8$. Minimizing the irritation in Cases I and II gives us $T_1 = 5.4$ and $T_1 = 3.7$ respectively. For the evening rush hour, minimizing the total waiting time and the irritation in Case I and II, leads to $T_1  = 13.3$, $T_1  = 14.0$ and $T_1 = 15.0$ respectively, see Figure \ref{fig:5}. They are different clearing strategies --  {\em a priori} it is not clear cut which one is the best.

\section{Discussion} \label{sec:4}


We used a discrete model to describe the cars traffic  in the presence of a traffic lights-controlled roadblock under heavy traffic flow conditions. Based on our model, we get the expressions (\ref{s1}) and (\ref{s2})    which are equivalent to those obtained in \cite{OPTI01} for the two one-way crossroads example. This indicates that the two one-way crossroad and the roadblock have a similar mathematical structure.\\
\\
For constant arrival and passing rates we obtain (\ref{eq:T_1}) as explicit expression for the green times at which the total waiting time is minimized. As main novelty, we introduce two  models for  the driver's irritation which finally lead to two clearing policies. The driver's irritation depends here linearly on the waiting time. At the first sight, this is reasonable because the longer the driver has to wait, the more irritated he or she will be. Such linear vs. nonlinear dependencies need to be further explored at a later stage, involving the input of group psychologists\footnote{We hope that group psychology is able to bring more behavioral insight in the modeling of (\ref{ir5}).}.
Here we assumed that the irritation per driver depends on the position of that driver in the queue at the moment the light switches from green to red. This can be explained in two distinct ways: The closer you are to the traffic light, the more irritated you become -- it was almost your turn to pass and now you have to wait again (Case I)! Alternatively, the closer you are to the traffic light, the less irritated you become -- now,  it is almost your turn to pass. The further you are from the traffic light the more irritated you are because you are wondering why the light switches (Case II). These modeling choices are based on how we experience irritation at traffic lights. There are many ways to model driver's irritation, but which is the right one? More research is needed for a good understanding of driver's irritation. Note that the numerical results reported in Section \ref{sec:3} indicate that minimizing the total waiting time looks quite similar to minimizing the irritation in Case I; the irritation in Case II is leading to somewhat different results.\\
\\
We applied our model to the traffic data received from Hillegom and found different values for the green times in the heavy traffic state when minimizing the total waiting time and the irritation. Therefore our model raises the natural question:

{\em What is the best strategy: maximize traffic throughput or minimize total driver's irritation?}

It is worth mentioning that, unlike \cite{KS} (the Kumar-Seidman network for production lanes) and \cite{Rooda}, e.g., our model is unable to preserve bounded waiting queues at roadblocks, and therefore can be classified as {\em unstable} (cf. \cite{SEL03}, e.g.).
We hope to be able to repair this within a modified framework: reformulate the roadblock problem in terms of a discrete particle system, where the irritation per particle (car) will be designed to depend also on the neighbors positions and velocities, which is at the moment not sufficiently well captured by ({\ref{ir5}); see \cite{Fermo}, e.g.  for a concrete possible direction. Such a modeling strategy is remotely resembling the stochastic dynamics of active Brownian particles.
As a next step, we intend also to apply this modeling framework to optimizing for load levels within capacity and will continue with further exploration  of the dependence of optimization on modeling of the concept of ``driver's irritation''.

\section{Acknowledgments} T.I.S. thanks Miguel Pinto for fruitful discussions on this topic. A.M. acknowedges discussions with Petre Cur\c seu (Tilburg University, The Netherlands) regarding a possible sociologic testing of our concept of irritation.

\bibliographystyle{plain}
\bibliography{CVleugels}

\end{document}